\documentstyle [11pt]{article}
\newcounter{zaehler}  
\newcommand{\al}{\alpha} \newcommand{\la}{\lambda}
\newcommand{\cm}{{\cal M}} \newcommand{\cn}{{\cal N}} \newcommand{\ck}{{\cal
K}}

\begin{document}
\setlength\baselineskip{24pt}

\title{\large\bf CELLULAR AUTOMATON MODEL OF PRECIPITATION/DISSOLUTION \
                  COUPLED WITH SOLUTE TRANSPORT}
\author{T. Karapiperis \\ Paul Scherrer Institute, CH-5232 Villigen PSI,
Switzerland}
\date{ }
\maketitle

\begin{abstract}
\indent

Precipitation/dissolution reactions coupled with solute transport are modelled
as a cellular automaton in which solute molecules perform a random walk on a
regular lattice and react according to a local probabilistic rule. Stationary
solid particles dissolve with a certain probability and, provided solid is
already present or the solution is saturated, solute particles have a
probability to precipitate. In our simulation of the dissolution of a solid
block inside uniformly flowing water we obtain solid precipitation downstream
from the original solid edge, in contrast to the standard reaction-transport
equations. The observed effect is the result of fluctuations in solute density
and diminishes when we average over a larger ensemble. The additional
precipitation of solid is accompanied by a substantial reduction in the
relatively small solute concentration. The model is appropriate for the
study of the r\^ole of intrinsic fluctuations in the presence of reaction
thresholds and can be employed to investigate porosity changes associated with
the carbonation of cement.

\end{abstract}

\setlength\baselineskip{18pt}

\section{Introduction}
\indent

Natural phenomena taking place in geological media often involve intricately
coupled physical and chemical processes. The dissolution of minerals by flowing
groundwater and their precipitation out of saturated solutions play a
fundamental r\^ole in the diagenetic changes of sediments and the weathering of
rocks. Precipitation/dissolution reactions can change rock properties (e.g.\
porosity, sorption capacity) and thereby influence the transport of solutes
through the rocks. It is clear that the assessment of natural and engineered
geological barriers aimed at containing the migration of soluble contaminants
will depend substantially on the coupling between solute-mineral reactions and
the water-conducting properties of the solid matrix. The reactions between
species in different phases are characterised by spatial and temporal
variations
in the set of reacting species, while the transfer of matter between the phases
results in moving solid boundaries. Depending on the time scales of interest, a
kinetic description of heterogeneous chemical reactions may become necessary
when the reaction rates are lower than the rates at which mobile reactants are
transported to the solid surfaces~\cite{w}. Although the problem of modelling
such systems can be addressed in principle in the framework of coupled
non-linear partial differential equations (PDE's), the numerical task is
usually
so complex (the numerical algorithm has to be continually readjusted to a
varying set of reactions and boundaries) that drastic approximations become
indispensable in practice \cite{l}. With a view to overcome at least some of
the
difficulties, we propose to model heterogeneous reactions coupled with mass
transport as a {\em cellular automaton (CA)}, taking advantage of the local and
parallel character of the natural processes.

The model described here continues and enriches the cellular automaton model of
coupled chemical reactions and mass transport that was presented in
Ref.~\cite{kb}. In that case, the time evolution of single-phase systems
involving particle transport (advection and diffusion/dispersion) coupled with
arbitrary chemical reactions was modelled by synchronous application, in
regular
time intervals, of a local probabilistic rule to all sites of a regular spatial
lattice spanning the extent of the physical system of interest. The model was
versatile enough to describe a variety of physical phenomena ranging from
fluctuation-dominated behaviour in the annihilation reaction $a + b \rightarrow
nothing$ to pattern formation with complex autocatalytic reaction schemes
\cite{kb}. The model was found capable of approximating the solution of PDE's,
while being able to account for microscopic effects that are typically washed
out by the averaging procedure that leads to the macroscopic equations. It
should be clear, however, that we do not aim at a strictly {\em microscopic}
description of physical phenomena and the `particles' in our simulations do not
correspond to individual molecules of the actual system. Accounting for such
inessential and hardly comprehensible detail would amount to an enormous waste
of resources. Instead, our {\em mesoscopic} approach accounts for those
elements
of microscopic reality (e.g.\ statistical fluctuations) that are likely to play
a r\^ole at the macroscopic level; it further models physicochemical processes
at a sufficiently elementary level to make the implementation of various
reaction schemes and boundary conditions, as well as their local readjustment
as
the system evolves, intuitively transparent.

When it comes to numerical implementation, the algorithms obtained in our
approach iterate an integer field (numbers of particles), which is free of
round-off errors, with real numbers appearing only in the form of
bounded probabilities. As a result, these algorithms are inherently stable.
By contrast, given the highly non-linear nature of the problems we wish to
solve, the stability of the respective finite difference equations (FDE's)
would
have to be tediously established on a case by case basis. At the same time,
however, the large size of the ensembles that have to be simulated in order to
achieve good statistics tends to slow down CA simulations. Thus, it became
clear
from the applications to single-phase systems that, for simple boundary
conditions and physicochemical parameters invariant in space and time,
conventional numerical techniques (e.g.\ FDE's) are computationally more
efficient than CA simulations. In such cases conventional methods for the
solution of the macroscopic equations are sufficient, unless they either suffer
from intractable stability problems or the impact of microscopic effects at the
macroscopic level cannot be neglected. Of course CA algorithms map naturally
onto the architecture of massively parallel computers, which are still in an
early stage of their development. We believe that future computer designs will
allow us to perform parallel simulations of large ensembles much more
efficiently and to thus better utilise the potential of CA algorithms.

The situation is, however, different in the presence of heterogeneous
reactions.
Precipitation or dissolution can only occur if solid is already present or the
solution is saturated. The existence of thresholds for the onset of the
reactions implies that the set of reactants may vary in space and time,
depending on whether certain minerals are reacting or not. Since the removal or
addition of solid changes the boundaries of the conducting channels, transport
properties are further closely coupled with reaction kinetics and may vary in
space and time. We believe CA modelling to be well-suited for modelling
heterogeneous reactions coupled with mass transport. A CA model can account for
spatial variations in flow and reaction parameters at no additional cost and,
by
modelling processes at an elementary level, it is capable of readjusting
locally
their characteristics as they vary in the process of the simulation. The
mesoscopic approach can be applied to solute-mineral reactions at different
spatial scales. On the one hand, one can address the problem at pore level
\cite{wjt}, by modelling the flow of water through the actual pores and
deriving
explicitly the time evolution of porosity and permeability. Alternatively, one
can model the system at a laboratory or field scale, assuming a
phenomenological
relation between the amount of precipitated or dissolved solid and the
associated porosity changes. In the latter case, which is the point of view
represented in this paper, one exploits the ease with which a CA models
processes with characteristics varying in space and time. These two levels of
modelling are to be seen as complementary, the simulations at pore scale
delivering the fundamental justification for the phenomenological relations
assumed at the larger scale.

This paper is organised as follows: The model is introduced in the following
section and numerical simulations are presented and discussed in a subsequent
section. Some theoretical questions are raised and future applications of the
model are proposed in a final section.

\section{Model of precipitation/dissolution}
\setcounter{equation}{0}
\indent

In our model physical and chemical processes take place on a regular spatial
lattice of spacing $\la$, according to a local rule applied synchronously to
all sites in regular time intervals $\tau$. The transport of solutes is
modelled
as a random walk, whereby particles move to neighbouring sites or remain at
their present site with prescribed probabilities. The macroscopic manifestation
of the random walk is a combination of {\em advection} and {\em diffusion}. The
displacement probabilities are chosen so that the desired {\em advection
velocity V} and {\em diffusion coefficient D} are obtained in the continuum
limit ($\la \rightarrow 0$ and $\tau \rightarrow 0$) \footnote{We assume that
{\em mechanical dispersion} can be macroscopically described by an effective
diffusion term, so `diffusion' should be understood in a broader sense than
just
{\em molecular diffusion}.}. Considering, for the sake of simplicity, a
one-dimensional system, we define the probabilities that a particle of solute
$s_\al$ moves to the next site on the right ($p_\al$) or the left ($q_\al$) so
that
\begin{equation}
V_\al \equiv (p_\al-q_\al){\la \over \tau}      \hspace{1cm} , \hspace{1cm}
D_\al \equiv (p_\al+q_\al){\la^2 \over 2\tau} \hspace{2.5mm} . \label{VD}
\end{equation}
In Eq.~(\ref{VD}), $V$ and $D$ are assigned a species label so that species
with
different transport properties can be modelled on the same lattice by choosing
$p_\al$ and $q_\al$ individually for each species. When taking the continuum
limit, we also let $p_\al - q_\al \rightarrow 0$ and we keep $\la^2 / 2 \tau$
and $(p_\al - q_\al)\la / \tau$ finite. Spatially varying $V_\al$ and $D_\al$
can be treated by making $p_\al$ and $q_\al$ position dependent; defining
$V_\al$ and $D_\al$ as above, we then identify the advection velocity with
$W_\al(x) \equiv V_\al(x) - dD_\al(x)/dx$.

Ref.~\cite{kb} contains a thorough discussion of our model of chemical
reactions
among solutes. Here we summarise the points that are relevant for the extension
to heterogeneous reactions. In solution we allow arbitrary reactions of the
type
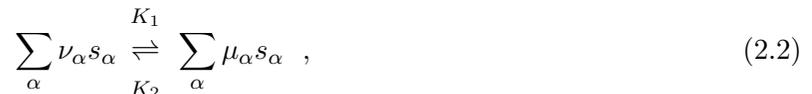
\begin{equation}
\sum_\al \nu_\al s_\al
\begin{array}{c} \mbox{\scriptsize $K_1$} \\ \rightleftharpoons \\
\mbox{\scriptsize $K_2$} \end{array}
\sum_\al \mu_\al s_\al \hspace{2.5mm} ,
\label{typ-rea}
\end{equation}
where $K_1$, $K_2$ are the {\em rate constants} and the {\em stoichiometric
coefficients} $\nu_\al$, $\mu_\al$ are non-negative integers. Referring for
brevity to a typical lattice site located at $x$ as `site $x$', we define the
{\em occupation number} $N_\al(x,t)$ as the number of particles of species
$s_\al$ that are to be found on site $x$, at time $t$. The action of the
chemical
reaction operator $\cal C$ on $N_\al(x,t)$ is defined by the equation
\begin{equation}
{\cal C} N_\al(x,t) = N_\al(x,t) + \sum_{r=1}^R
\left( \nu_{\al r}^{(f)} - \nu_{\al r}^{(i)} \right) \eta_{x,r} \hspace{2.5mm}
,
\label{reac-op}
\end{equation}
where $\eta_{x,r}$ is a random Boolean variable and $\nu_{\al r}^{(f)}$ and
$\nu_{\al r}^{(i)}$ are stoichiometric coefficients referring to products
and reactants, respectively. The summation runs over all $R$ one-way reactions
obtained by separating the forward from the reverse process (irreversible
reactions are added to the list as they are). Reaction $r$ takes place when
$\eta_{x,r} = 1$, which happens with probability
\begin{equation}
\wp (\eta_{x,r}=1 | \{ N_\beta(x,t): s_\beta \in {\cal S} \} ) \equiv
P_r F_r \left( \left\{ N_\beta(x,t) : s_\beta \in {\cal S} \right\} \right)
\hspace{2.5mm} , \label{reac-prob}
\end{equation}
where $P_r$ is a real constant, $F_r$ is a function of the occupation numbers
and $\cal S$ is the set of all species in solution. According to chemical
reaction rule II of Ref.~\cite{kb},
\begin{equation}
F_r \left( \left\{ N_\beta(x,t): s_\beta \in {\cal S} \right\} \right) =
\prod_\beta\prod_{m=1}^{\nu_{\beta r}^{(i)}} \left( N_\beta(x,t) - m + 1
\right)
\hspace{2.5mm} .      \label{F2}
\end{equation}
For example, if $r$ is the reaction $a + b \rightarrow c$, then it occurs with
probability $P_r N_a N_b$. In deriving the continuum limit of the full model
we also take $P_r \rightarrow 0$, keeping $P_r / \tau$ finite. Of course, care
has to be taken so that the probability defined by Eqs.~(\ref{reac-prob}) and
(\ref{F2}) does not exceed 1. It is always possible to ensure this by choosing
$P_r$ sufficiently small (so that $P_r N_a N_b \leq 1$, $\forall x,t$ in the
above example), since the occupation number remains finite during a simulation.

The particle density $\rho_\al(x,t)$ is defined by averaging the occupation
number $N_\al(x,t)$ over an ensemble of macroscopically identical systems.
In Ref.~\cite{kb} it is shown how rule (\ref{F2}) leads, under the assumptions
of {\em molecular chaos} (no correlations between the occupation numbers of
different species) and a smooth particle density function, to the standard
rate law, which is consistent with the {\em law of mass action} at equilibrium.
The rate constant $k_r$ is given by
\begin{equation}
k_r \equiv {P_r \over \tau} \hspace{2.5mm} . \label{kr}
\end{equation}
Thus, for the reaction $a + b \rightarrow c$ we obtain a rate of $k_r \rho_a
\rho_b$.

The novel element in solute-mineral reactions consists in the existence of a
precipitation threshold: solid must already be present or the solution must be
saturated before the reaction may proceed. We investigate the simple
precipitation/dissolution reaction
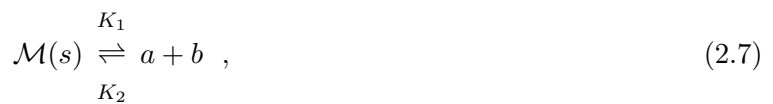
\begin{equation}
{\cm}(s) \begin{array}{c} \mbox{\scriptsize $K_1$} \\ \rightleftharpoons \\
\mbox{\scriptsize $K_2$} \end{array} a + b \hspace{0.25cm} , \label{pre-dis}
\end{equation}
where $a$, $b$ are aqueous species, $\cm$ is a solid and $K_1$, $K_2$ are the
rate constants. We assume $\cm$ to be a mineral occupying a section of an
otherwise inert rock of uniform porosity $\epsilon$. Water flows through the
rock and gradually dissolves the mineral according to the forward reaction in
(\ref{pre-dis}). The solutes produced are then transported by advection,
mechanical dispersion and molecular diffusion; in the process they may
recombine
and precipitate according to the reverse reaction in (\ref{pre-dis}). In this
paper we treat the porosity as a constant, making sure that the amount of
dissolved and reprecipitated mineral is small enough to make a readjustment of
the porosity unnecessary. This is clearly a restriction that we can easily
relax
when we address the question of porosity changes caused by solute-mineral
reactions. At this stage, however, a variable porosity would unnecessarily
complicate our effort to compare our model to the standard macroscopic
approach.

The macroscopic reaction-transport equations for the system considered here are
\begin{eqnarray}
\epsilon {\partial C_a(x,t) \over \partial t} & = &
- \epsilon V_a \frac{\partial C_a(x,t)}{\partial x}
+ \epsilon D_a \frac{\partial^2 C_a(x,t)}{\partial x^2}
+ \zeta(x,t) \left( K_1 - \epsilon^2 K_2 C_a(x,t) C_b(x,t) \right) \nonumber \\
{\partial C_\cm(x,t) \over \partial t} & = &
- \zeta(x,t) \left( K_1 - \epsilon^2 K_2 C_a(x,t) C_b(x,t) \right)
\hspace{2.5mm} ,
\label{pre-dis-eq}
\end{eqnarray}
where $C_a$ and $C_b$ are the concentrations of the solutes in $mol$ per $m^3$
of water and $C_\cm$ is the `concentration' of solid in $mol$ per $m^3$ of
rock. $\zeta$ takes the value 0 or 1 according to
\begin{equation}
\zeta(x,t) =
\begin{array}{cl} 1 \hspace{1mm} , & \hspace{2mm} \mbox{if} \hspace{2mm}
C_\cm(x,t) > 0 \hspace{2mm} \mbox{or} \hspace{2mm} C_a(x,t) C_b(x,t) > \ck \\
0 \hspace{1mm} , & \hspace{2mm} \mbox{otherwise} \end{array}
\hspace{0.25cm} ,
\end{equation}
where $\ck \equiv K_1 / \epsilon^2 K_2$ is the {\em solubility product}. The
equation for $C_b$ is analogous to the one for $C_a$.

We model the transport of the mobile species $a$ and $b$ as a random walk on a
lattice, as explained in the beginning of this section. The transport operation
leaves $\cm$-particles unaffected. After the transport operation has been
applied synchronously to the entire lattice, the evolution of the system from
time $t$ to $t+\tau$ is completed by application of the chemical reaction
operation. The local rule for the reaction (\ref{pre-dis}) is defined in terms
of two constants, $P_1$ and $P_2$, and two random Boolean variables,
$\eta_{x,1}$ and $\eta_{x,2}$. $\eta_{x,1} = 1$ ($\eta_{x,2} = 1$) signals the
occurrence of a dissolution (precipitation) event at $x$.

In dissolution, solid $\cm$-particles are converted to solute $a$- and
$b$-particles. Dissolution is of course possible at a site $x$ if one or more
$\cm$-particles are present there. If this condition is fulfilled, then we
choose $\eta_{x,1} = 1$ with probability $P_1$, which is independent of the
amount of solid available. If $\eta_{x,1} = 1$, the number of $\cm$-particles
at
$x$ is diminished by 1, while those of $a$- and $b$-particles are increased
by~1. At most one $\cm$-particle may dissolve per site during an evolution step
from $t$ to $t+\tau$.

Precipitation results in the replacement of solute $a$- and $b$-particles by
solid $\cm$-particles. The process is possible at a site $x$ if at least one
$\cm$-particle is already present there or if the product of the densities of
$a$ and $b$ (calculated as ensemble averages of the respective occupation
numbers $N_a$ and $N_b$) is greater than $P_1 / P_2$. In the former case
precipitation occurs around a nucleus of already present solid, whereas the
latter inequality amounts to the saturation condition of the solution. If at
least one of these conditions is fulfilled, then we choose $\eta_{x,2} = 1$
with
probability $P_2 N_a N_b$. We note that the saturation condition is the same
for
the entire ensemble, whereas the reaction probability is defined individually
for each member. The probability of precipitation is naturally higher the
greater the number of solute particles present and vanishes when either or both
species are absent. If $\eta_{x,2} = 1$, the number of $\cm$-particles at $x$
is
increased by 1, while those of $a$- and $b$-particles are diminished by 1. At
most one $a$- and one $b$-particle may be subtracted per site during an
evolution step. The definition of the reaction probability for precipitation
guarantees that there is a sufficient amount of solute particles when the
reaction proceeds.

Referring to the forward and backward reactions in (\ref{pre-dis}) as reactions
1 and 2, respectively, we write the probability of their occurrence as
\begin{eqnarray}
\addtocounter{zaehler}{1}
\hspace*{-1.5cm} \wp ( \eta_{x,1} = 1 ) & \equiv & P_1
\theta \left( N_\cm(x,t) \right) \hspace{2.5mm} , \\
\addtocounter{equation}{-1} \addtocounter{zaehler}{1}
\hspace*{-1.5cm} \wp ( \eta_{x,2} = 1 ) & \equiv & P_2 N_a(x,t) N_b(x,t)
\left[ \theta \left( N_\cm(x,t) \right) + \delta_{N_\cm(x,t),0} \;
\theta \! \left( \rho_a(x,t) \rho_b(x,t) - {P_1 \over P_2} \right) \right]
\hspace{2.5mm} . \label{pre-dis-rule}
\end{eqnarray}
We define $\rho_\al(x,t) \equiv < N_\al(x,t) >$, where $<\ldots>$ denotes an
ensemble average. The value of the $\theta$-function $\theta(z)$ is 1 if $z >
0$
and 0 otherwise. $\delta_{i,j} = 1$ or 0 according as the integers $i$ and
$j$ are equal or not.

Assuming no correlations between $a$-, $b$- and $\cm$-particles (molecular
chaos), we obtain the reaction rate
\begin{eqnarray}
\setcounter{zaehler}{0}
{\partial \rho_\cm(x,t) \over \partial t} & = & \mbox{} - {P_1 \over \tau} \wp
\left( N_\cm(x,t) > 0 \right) + {P_2 \over \tau} \rho_a(x,t) \rho_b(x,t) \cdot
\nonumber \\
& & \left[ \wp \left( N_\cm(x,t) > 0 \right) + \wp \left( N_\cm(x,t) = 0
\right)
\; \wp \! \left( \rho_a(x,t) \rho_b(x,t) > {P_1 \over P_2} \right) \right]
\hspace{2.5mm} . \label{pre-dis-rate}
\end{eqnarray}
The same term, but with opposite sign, appears on the RHS of the
reaction-transport equations derived in the continuum limit for $a$ and $b$.
$\wp \left( N_\cm(x,t) = 0 \right)$ denotes the probability that there are no
solid particles at the given location and time, and $\wp \left( N_\al(x,t) > 0
\right) = 1 - \wp \left( N_\al(x,t) = 0 \right)$. $\wp \left( \rho_a(x,t)
\rho_b(x,t) > P_1 / P_2 \right)$ is the probability that the saturation
condition is fulfilled. For a given spatially smooth density $\rho_\al(x,t)$ of
particles we expect these probabilities to depend on the macroscopic parameters
of the system. Thus, for {\em randomly moving} particles, such as the solutes
in
our model, the occupation numbers obey the Poisson distribution $\wp \left(
N_\al(x,t) = n \right) = e^{-\rho_\al(x,t)} \rho_\al^n(x,t) / n! \, , \; n \geq
0$. As we shall see in the next section, this is not true about the immobile
solid particles.

The transition from particle densities to concentrations is made by
the relations $C_a=\gamma \rho_a$, $C_b = \gamma \rho_b$ and $C_\cm =
\gamma \epsilon \rho_\cm$, where the constant $\gamma$ can be fixed by
relating the initial concentrations of the physical problem to the initial
particle densities of the simulation. The macroscopic equations obtained from
our mesoscopic model differ from Eqs.~(\ref{pre-dis-eq}) in the precise
definition of the switch that signals the onset of precipitation/dissolution,
but this difference is only likely to play a r\^ole for very small solid
particle densities. Comparing with (\ref{pre-dis-eq}), we are led to relate the
rate constants to the reaction parameters used in the simulation by $K_1 =
\gamma\epsilon P_1/\tau $ and $K_2 = P_2 / \gamma\epsilon\tau$.

\section{Simulations of precipitation/dissolution}
\setcounter{equation}{0}
\indent

We have applied the CA model described in the previous section to systems with
one-dimensional geometry, assuming translational invariance in the other two
dimensions. In this section we compare the results of our simulations with the
solution of the macroscopic reaction-transport equations.

We consider first the dissolution of a small amount of solid ($\epsilon \simeq
1$) placed at $t = 0$ in a stream of uniformly flowing water. Solid and solute
particles reside on a one-dimensional lattice of $\cn_x = 101$ sites and
spacing
$\la$. The water is not explicitly modelled. Initially we place randomly 510
particles of the solid $\cm$ on the left half of the otherwise empty lattice,
to
an average density of 10 particles/site. The solid particles do not move. Solid
can, however, be transferred across the lattice indirectly by dissolution,
transport of the solutes and reprecipitation. At each site where solid is
present there is a probability of $P_1 = 0.04$ that a solid particle dissolves
during an update. Dissolution amounts to disappearance of an $\cm$-particle and
creation of one $a$- and one $b$-particle in solution. The solutes $a$ and $b$
are carried along by flowing water with an advection velocity of $V_a = V_b =
0.1 \, \la/\tau$ (i.e.\ it would take a solute particle 1000 updates, applied
in
time intervals of $\tau$, to traverse the lattice by advection alone) from left
to right. The solutes are additionally subject to diffusion/dispersion with a
diffusion coefficient of $D_a = D_b = \la^2/2\tau$. In the simulation, solute
particles perform a random walk with displacement probabilities of $p_a = p_b =
0.55$ to the right and $q_a = q_b = 0.45$ to the left. At every time step we
evaluate the particle densities $\rho_a(x,t)$ and $\rho_b(x,t)$ by averaging
the
particle numbers $N_a(x,t)$ and $N_b(x,t)$ over an ensemble of $\cn_y$ systems.
One can think of the ensemble members as being aligned in a $y$-direction
orthogonal to the actual $x$-direction of the system; then the averaging is
performed over $y$, for fixed $x$ and $t$. Defining the parameter $P_2 = 0.4$,
we specify that precipitation may take place only if $N_\cm(x,t) > 0$ or
$\rho_a(x,t) \rho_b(x,t) > P_1/P_2$. Once allowed, precipitation, i.e.\ the
reverse of the dissolution process described above, proceeds with probability
$P_2 N_a(x,t) N_b(x,t)$.

Solute particles are subject to the following boundary conditions: The boundary
site on the left ($x = 0$) is a sink absorbing all $a$- or $b$-particles that
reach it. At the right boundary, $x = L \equiv (\cn_x-1) \la$, we impose the
macroscopic boundary condition $\left. \partial\rho_a(x,t) / \partial x
\right|_{x=L} = \left. \partial \rho_b(x,t) / \partial x \right|_{x=L} = 0$.
Equivalently, we set the diffusive flux, which is proportional to the density
gradient, equal to zero at $x = L$; this leaves only advection to take care of
net solute transport across the boundary. To implement the zero-gradient
boundary condition in our model we follow Ref.~\cite{kb}: the lattice is
formally extended by one lattice spacing; before each transport operation the
occupation number of the new site ($x = L+\la$) is set equal to that of the
site immediately preceding the boundary ($x = L-\la$). As long as the particle
density is not the same at $L$ and $L-\la$, a net amount of particles will
diffuse either from $L$ to $L-\la$ and $L+\la$ or vice versa until there is no
density gradient at the boundary. In physical terms, we superpose equal and
opposite amounts of outgoing and incoming diffusive flux and thus make the
total
diffusive flux vanish.

Fig.~\ref{fig-comp} displays the profile of solid particle density obtained
from
the above simulation after $\cn_t = 5000$ iterations. The solid, dashed and
dotted curves show the results obtained from three different simulations with
ensemble sizes $\cn_y = 250$, 1000 and 4000, respectively. The dot-dashed curve
shows the result of the macroscopic transport equations (obtained from
(\ref{pre-dis-eq}) by setting $\epsilon$, $\gamma$, $\la$ and $\tau$ equal to
unity). We see that a large fraction of the original solid has dissolved. In
the
simulations, solid reprecipitates downstream from the original solid edge, in
contrast to the macroscopic description where the edge remains sharp. The
observed effect is entirely due to statistical fluctuations in the solute
densities: due to fluctuations, the product of the densities occasionally
exceeds the solubility limit and precipitation ensues. Fig.~\ref{fig-comp}
shows
that averaging over a larger ensemble reduces the fluctuations and the
resulting
precipitation.

\begin{figure}[t]
  \vspace{9.0cm}
  \caption{Density profile of solid particles. The meaning of the different
  curves is explained in the text. \label{fig-comp} }
\end{figure}

The difference between the simulations and the macroscopic equations on the
left
side of the solid block is of different origin:

$(i) \;$ It does not depend on the size of the ensemble and is therefore not of
statistical origin.

\begin{figure}[t]
  \vspace{9.0cm}
  \caption{Distribution of solid-particle occupation numbers at $x = 40 \la$
and
  (a) $t=0$, (b) $t = 5000 \tau$. The meaning of the different symbols and
  curves is explained in the text. \label{fig-dist} }
\end{figure}

$(ii) \;$ It depends weakly on the number of transport steps per reaction step,
$n_D$. Increasing $n_D$ washes out reaction correlations between $a$- and
$b$-particles; in Ref.~\cite{kb}, such correlations were found to influence the
macroscopic properties of a system of diffusing $a$-, $b$- and $c$-particles
subject to the reaction $a + b \rightleftharpoons c$; it was also found that
correlations were practically eliminated for $n_D = 10$ (e.g.\ the remaining
effect on the equilibrium reaction quotient $\rho_c / \rho_a \rho_b$ was less
than $0.3 \%$ for a homogeneous system). For the precipitation/dissolution
reaction we find that increasing $n_D$ from 1 to 4 (while keeping $V_\al$ and
$D_\al$ fixed by appropriate readjustment of $p_\al$ and $q_\al$) reduces the
solid particle density by 20 - 30 $\%$ in the middle of the upstream tail of
the
density profile, but increasing $n_D$ further to 9 makes only a 2 - 3 $\%$
difference.

$(iii) \;$ It diminishes systematically as we approach the continuum limit by
refining the discretisation of the lattice (i.e.\ increasing $\cn_x$ while
keeping macroscopic parameters fixed by appropriate readjustment of $\cn_t$,
$p_\al$, $q_\al$ and $P_r$). We have confirmed this for $\cn_x = 200$ and 400,
but were prevented from implementing larger lattices by CPU time limitations.
It
is worth noting that the amount of solid precipitating downstream is left
unaffected (within fluctuations) by the refined discretisation for given
ensemble size.

We next look at a cross-section of the ensemble of systems along the fictitious
$y$-direction: for given $x$ and $t$, there is an interesting difference
between
the distributions of solute and solid occupation numbers within the ensemble.
For the solutes $a$ and $b$, occupation numbers are distributed, within the
statistics of the ensemble, according to the Poisson distribution consistent
with the average density. This is expected due to the randomising action of the
motion of these particles. Since solid particles are placed randomly, as
described above, their initial distribution is also Poisson \footnote{The
macroscopic fate of the system would be the same if exactly 10 solid particles
were placed at every site initially.}. The initial distribution of occupation
numbers at $x = 40\la$ is shown in Fig.~\ref{fig-dist}a for the three
simulations considered in Fig.~\ref{fig-comp} (solid circles, open circles
and crosses, respectively); for comparison, the solid, dashed and dotted curves
show the Poisson distributions for the respective particle densities. A similar
plot at time $t = 5000 \tau$ (Fig.~\ref{fig-dist}b) reveals a much wider
distribution of occupation numbers. In this case we observe a remarkably high
frequency of sites with no solid particles. This is related to the small
initial
particle density. In simulations with 10 and 100 times higher initial
densities,
the frequency of empty sites is very small ($\sim 1 \%$) and negligible
respectively, but occupation numbers still follow broad distributions.

\addtocounter{zaehler}{1}
\begin{figure}[t]
  \vspace{9.0cm}
  \caption{Profile of solid `concentration' (solid line: simulation, dotted
  line: macroscopic equations). \label{fig-calc-a} }
\addtocounter{figure}{-1} \addtocounter{zaehler}{1}
  \vspace{9.0cm}
  \caption{Concentration profiles of solute (horizontal dotted line and solid
  line running on the average parallel to it: macroscopic equations and
  simulation, respectively; dashed line is a smoothed version of solid line;
  axis labelled on the right) and solid (vertical dotted curve and solid line
  running partly parallel to it: macroscopic equations and simulation,
  respectively; axis labelled on the left). \label{fig-calc-b} }
\end{figure}
\setcounter{zaehler}{0}

In order to appreciate the implications of density fluctuations for models of
the dissolution of actual minerals, we consider now a piece of rock of length
$L = 10 m$, with a uniform porosity of $\epsilon = 11 \%$. The left half of the
rock is occupied at
time $t=0$ by a soluble mineral with a density of 2.71 $g/cm^3$ and a molecular
weight of 100. The rate constants for dissolution and precipitation are $K_1 =
1.2 \times 10^{-9} mol \; m^{-3} s^{-1}$ and $K_2 = 3.3 \times 10^{-5} m^3
mol^{-1} s^{-1}$, respectively. Thus, the solubility product is $\ck =
10^{-2.5} mol^2 m^{-6} = 10^{-8.5} (mol/\ell)^2$. The physical and
(equilibrium)
chemical parameters are chosen to be consistent with calcite ($CaCO_3$) as a
typical mineral in the geological systems of interest. The dissolved species
$Ca^{2+}$ and $CO_3^{2-}$ move with an advection velocity of $10^{-8} m/s$ and
a diffusion coefficient of $10^{-9} m^2/s$.

For the simulation we choose $\la = 10^{-2} m$, $\tau = 5 \times 10^4 s$ and
$\gamma = 0.28 \, mol/m^3$. At time $t = 10^{10} s$ a small fraction of the
solid has been depleted from the left side of the original block of mineral
(Fig~\ref{fig-calc-a}). The simulation agrees well with the result of the
macroscopic equations (\ref{pre-dis-eq}); the remaining small discrepancy is
qualitatively reminiscent of the relatively larger one observed in
Fig.~\ref{fig-comp}. Solid precipitates downstream from the original solid
edge, an effect which is absent from
\linebreak
\clearpage
\noindent
the result of the macroscopic equations
(Fig~\ref{fig-calc-b}). The amount of precipitating solid diminishes as the
size
of the ensemble increases (not shown in the figure, which was obtained with
$\cn_y = 200$),
pointing, as already
remarked in connection with Fig.~\ref{fig-comp}, to the crucial r\^ole of
statistical fluctuations. The amount of solid that is found downstream from the
edge is tiny on the scale of the total amount of solid, but the consequences
can be more serious for the concentration of solutes. The net precipitation of
solid in the right half of the lattice is accompanied by a reduction in solute
concentration relative to the prediction of the macroscopic equations. In
Figs.~\ref{fig-calc-a} and \ref{fig-calc-b}, of 400 million solid particles
initially present, about 4500 have been removed from the solid block and about
500 have precipitated in the right half of the lattice. This suffices, however,
to produce a 20-25~$\%$ suppression in solute concentration, as a comparison of
the dotted and dashed curves shows. The importance of this suppression will
depend of course on the particular situation, but it illustrates the fact that
corrections to reaction rates can have a very different impact on species whose
concentrations differ by several orders of magnitude (5-6 in the case at hand).
The size of the effect on the spatial distribution of the mineral itself is
likely to become significant at sufficiently long times. When the time scale is
long enough for precipitation/dissolution to cause changes in the porosity,
statistical fluctuations in solute concentrations will certainly influence
those
changes, to a degree, however, that remains to be investigated.

\section{Discussion and outlook}
\indent

The mesoscopic model of precipitation/dissolution introduced here is seen as
part of a comprehensive model of the coupled transport and chemical reactions
of
solutes in geological media. As in the model of transport and reactions in
single-phase systems presented in Ref.~\cite{kb}, physical and chemical
processes are modelled as simple CA rules applied to a collection of particles
on a regular lattice. One can thus model spatial and temporal variations in the
characteristics of the system by local readjustment of parameters without
changing the overall algorithm. This, coupled with the guaranteed stability of
our algorithms, provides a significant advantage over standard numerical
methods
when modelling heterogeneous reactions. Solute-mineral reactions have the
distinctive feature that they become possible only beyond a certain threshold,
so that the set of reactions itself and the boundaries between different phases
may vary in space and time. Our model provides further a natural way to study
the r\^ole of fluctuations in coupled physicochemical processes. We have seen
one distinctive consequence of statistical fluctuations in solute
concentrations
in the form of mineral precipitation beyond solid edges that would remain sharp
according to the standard macroscopic equations. This has a significant effect
on the relatively (compared to the mineral) small concentrations of solutes,
but
it may also influence the porosity of the solid matrix.

Since the particles of our simulations do not model individual molecules of the
actual system, the interpretation of statistical fluctuations and their
consequences is not immediately obvious. We assume here a phenomenological
point
of view, allowing for microscopic processes not explicitly accounted for in our
model to contribute to the fluctuations we model. Then one has to fix the
amplitudes of the fluctuations by comparing the size of the resulting effects
with real systems. From a theoretical point of view our model of
precipitation/dissolution can be seen as a paradigm of a reaction-transport
system subject to reaction thresholds. It would be interesting to develop the
corresponding stochastic differential equations with the appropriate noise
amplitude and compare their results with those of our simulations. One could
thus study in a qualitatively different framework the r\^ole of intrinsic noise
along the lines of the work reported in Ref.~\cite{wdbs}.

In our simulations so far we have modelled the transport of solutes assuming
that their advection follows a given velocity field. If the solute
concentration
is small enough, the velocity field coincides with that of pure water flowing
under the given mechanical conditions. The dynamics of water flow is governed
by the Navier-Stokes equations, which for a porous medium go over to Darcy's
law
on a macroscopic scale. The problem of solute transport can be solved by a
two-step process, where one first solves the appropriate hydrodynamic equations
(by CA or other numerical methods) and then uses the resulting velocity field
as
an input parameter to the equations of motion of the solute. It may, however,
be
more economical to model the motion of the solute directly by incorporating a
random walk in the evolution rule of a {\em lattice gas}, i.e.\ one of the CA
used for the simulation of fluid dynamics~\cite{LGA}. Chemical reactions can be
added and the full algorithm is iterated in time taking into account possible
changes in the mechanical properties of the solid matrix due to the reactions.

The development of our CA model for precipitation/dissolution was motivated
largely by its intended application to the investigation of the evolution of
porosity as a result of solute-mineral reactions. A simple system which
possesses the necessary ingredients for such an investigation consists of a
block of cement ($Ca(OH)_2$) dissolving in the presence of $CO_3^{2-}$ (e.g.\
through contact with atmospheric air) and reprecipitating as $CaCO_3$. A simple
model of the carbonation of cement appeared previously in Ref.~\cite{bb}. The
chemistry of the system can be schematically represented by the reactions
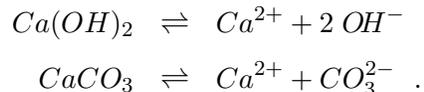
\begin{eqnarray*}
Ca(OH)_2 & \rightleftharpoons & Ca^{2+} + 2 \; O\!H^{-} \\
CaCO_3 & \rightleftharpoons & Ca^{2+} + CO_3^{2-} \hspace{2.5mm} .
\end{eqnarray*}
Significant porosity changes by $CaCO_3$ precipitation have been observed
experimentally \cite{sbps}. They can be crucial, for example, in the assessment
of cement as a barrier to the migration of soluble contaminants released
following a failure in a waste repository. Our approach offers the possibility
of developing flexible models of heterogeneous reactions coupled with mass
transport and reactions in the aqueous phase. Applications to real geological
systems will help us appreciate better the r\^ole of intrinsic fluctuations.
By developing in parallel the corresponding stochastic differential equations
and comparing them with mesoscopic simulations, we can test to what extent the
full detail of the latter can be effectively grasped in an more economical
modelling framework. The relative computational efficiency of the different
approaches will depend largely on the specific applications, as well as on
future developments in massively parallel architectures and hardware.

\subsection*{Acknowledgements}
\indent

The author has benefited a great deal from discussions with U. Berner, J.
Hadermann and F. Neall. Partial financial support by NAGRA is gratefully
acknowledged.

\end{document}